\documentstyle[preprint,pra,aps]{revtex}

\begin{document}

\draft   

\title{General fidelity limit for quantum channels}

\author{Howard Barnum, Christopher A.~Fuchs,\cite{FuchsAddress}
Richard Jozsa,\cite{JozsaAddress} and Benjamin
Schumacher\cite{SchumacherAddress}}
\address{Center for Advanced Studies, 
Department of Physics and Astronomy,\\
University of New Mexico, Albuquerque, New Mexico 87131--1156}

\date{5 March 1996}

\maketitle
 
\begin{abstract}
We derive a general limit on the fidelity of a quantum channel
conveying an ensemble of pure states.  Unlike previous results, this
limit applies to arbitrary coding and decoding schemes.  This
establishes the converse of the quantum noiseless coding theorem for
all such schemes.
\end{abstract}


\section{Quantum encoding and decoding}

One of the central problems in quantum information theory \cite{chb}
is the transmission of pure quantum states from a sender to a receiver
using the least possible channel resources.  Suppose Alice generates
the state $|a_{i}\rangle$ of the system $Q$ with probability $p_{i}$.
This is encoded by some (possibly mixed) state $W_{i}$ of the channel
system $C$ (generally of smaller Hilbert-space dimension than $Q$) and
delivered to Bob, who performs a decoding operation giving a state
$w_{i}$ of $Q$.  We assume that no ``noise'' is present in the system
except that introduced in the coding and decoding processes.  Letting
$\pi_{i} = |a_{i}\rangle\langle a_{i}|$, this may be represented by
\begin{displaymath}
        \pi_{i}  \longrightarrow
        W_{i}  \longrightarrow  w_{i} \;.
\end{displaymath}
The decoded state $w_{i}$ is not necessarily required to equal
$\pi_{i}$ exactly; it will suffice for Alice and Bob if the inputs and
outputs are sufficiently close to each other.  The ``closeness'' of
the input and output states is measured by the {\em average
fidelity\/} $\overline{F}$:
\begin{equation}
        \overline{F} = \sum_{i} p_{i} \, F(\pi_{i}, w_{i})\;,
\end{equation}
where $F(\pi_{i},w_{i}) = {\rm Tr}\, \pi_{i} w_{i}$ is the probability
that $w_{i}$ will pass a test that checks its identity against
$\pi_{i}$.  Alice and Bob will succeed in their task if $\overline{F}$
is close to unity, and fail if it is not.  Our problem is to
characterize the minimal channel resources, i.e., the minimal
dimension of the support of the states $W_i$, which are necessary and
sufficient for high fidelity transmission \cite{schu,js94}.

This process of retrieving faithful copies of the input states from
the states of the channel has applications in quantum cryptography,
where nonorthogonal states represent encrypted classical information
\cite{Bennett84,Bennett92}, and
in problems of efficient information storage and retrieval for quantum
computers \cite{DiVincenzo95}.

The decoding operation $W_{i} \rightarrow w_{i}$ must be accomplished
without any ``side information''---i.e., the only information
possessed by Bob about the input state is his knowledge of the message
ensemble and the coding procedure that prepares the channel $C$. Bob's
decoding procedure must be a dynamical evolution that is specified
apart from the state on which it acts.  On the other hand, we make no
such assumption about Alice's encoding operation, so that the
association $\pi_{i} \rightarrow W_{i}$ is completely arbitrary.
Indeed we generally allow Alice to have knowledge of the identities of
the specific input states and she is therefore able to effect
arbitrary encodings. In contrast, Bob is unable to reliably identify
the (generally nonorthogonal) channel states $W_i$
\cite{Wootters82,asher}
so his decoding procedure is restricted by the laws of quantum
mechanics as described in \S 3 below.

Note that the encoding procedure here is more general than the
scenario in which Alice is required to encode the input states {\em
without\/} knowledge of their identities (knowing only their a priori
distribution). In this situation the allowable encodings
$\pi_i\rightarrow W_i$ are no longer arbitrary but subject to
restrictions analogous to those on Bob's decoding procedures.  (This
is in contrast \cite{asher,GP} to the corresponding situation with
{\em classical\/} signals which may always be reliably identified
without disturbance.)  A remarkable consequence of the quantum
noiseless coding theorem and its converse described below is that the
minimal channel resources for high fidelity transmission in this
situation are asymptotically the {\em same\/} as those for the case
where Alice is able to apply arbitrary encoding processes, i.e.,
knowledge of the identity of the input states does not lead to any
reduction of channel resources. Indeed in
\cite{schu,js94} an explicit encoding scheme is described
which achieves (asymptotically) the minimal channel resources
and this scheme operates without knowledge of the identity
of the input states (being dependent only on their a priori
distribution).

The quantum noiseless coding theorem proved in \cite{schu,js94}
relates the achievable average fidelity $\overline{F}$ to the size of
the channel system.  This size is given in terms of the number of
two-level systems, or {\em qubits}, that comprise the channel when
coding is performed on large blocks of signals drawn identically from
the original message ensemble.\footnote{Of course, the description of
the channel in terms of qubits is mere convenience.  Any channel
described by a Hilbert space of dimension $d$ is equivalent for our
purposes to $\log d$ qubits.}  Suppose we have input states $\pi_{i}$
with probabilities $p_{i}$, as before, and let $\rho = \sum_{i} p_{i}
\pi_{i}$ be the density operator describing the input ensemble.  The
von Neumann entropy of $\rho$ is given by
\begin{equation}
        S(\rho) = - {\rm Tr}\, \rho \log \rho \;,
\end{equation}
where the base of the logarithm is 2.  Then the quantum noiseless
coding theorem states:
\begin{quote}
        Let $\epsilon, \delta > 0$, and suppose $S(\rho) + \delta$
        qubits are available in the channel per input state.  Then for
        all sufficiently large $N$, there exists a coding and a
        decoding scheme which transmits blocks of $N$ states with
        average fidelity $\overline{F} > 1 - \epsilon$.
\end{quote}
In other words, the von Neumann entropy is a measure of the channel
resources (in qubits) sufficient to transmit quantum states with
arbitrarily high average fidelity.  A converse to the theorem has also
been given.
\begin{quote}
        Let $\epsilon, \delta > 0$, and suppose $S(\rho) - \delta$
        qubits are available in the channel per input state.  Then for
        all sufficiently large $N$, for any coding and decoding scheme
        for blocks of $N$ states, the average fidelity satisfies
        $\overline{F}<\epsilon$.
\end{quote}
This converse states that the von Neumann entropy is a measure of the
channel resources {\em necessary\/} to transmit quantum states with
high average fidelity.

In this formulation, the converse refers to all possible
coding/decoding schemes.  However, the proof given in \cite{schu} and
\cite{js94} implicitly assumes that the decoding scheme is {\em
unitary}---that is, that the map $W_{i} \rightarrow w_{i}$ is a
unitary mapping from the channel's Hilbert space into the Hilbert
space of the decoded signals.  There are still other possibilities
that must be considered.  For example, the decoding scheme might
involve a measurement, the discarding of an entangled subsystem, or
any other process allowed within the the laws of physics.  The
converse of the quantum noiseless coding theorem cannot be established
in full generality without considering all conceivable decoding
schemes.  Indeed in an Appendix we present a simple example containing
all the salient features of this problem that shows for {\em
particular\/} (nonoptimal) encodings it is possible for nonunitary
decodings to provide higher fidelity than any unitary decoding scheme.
Therefore the issue of real concern for the converse is whether such
nonunitary decoding schemes add any power to optimal encodings.

Our aim in this paper is to complete the general proof of the converse
of the quantum noiseless coding theorem by establishing a lemma that
links the average fidelity $\overline{F}$ of the decoded signal states
to the size of the channel system and to properties of the density
operator $\rho$ of the ensemble of input states.  This fidelity lemma
may also prove useful in other contexts.

\section{Fidelity}

Suppose $\rho_{1}$ and $\rho_{2}$ are density operators describing
states of a quantum system $Q$.  We can always imagine that these
mixed states arise by a partial trace operation from pure states of an
extended system $QA$.  That is, there are states $|1\rangle$ and
$|2\rangle$, called ``purifications'' of $\rho_{1}$ and $\rho_{2}$,
for which
\begin{eqnarray*}
        \rho_{1} & = & {\rm Tr}_{A}\, |1\rangle\langle 1| \\ \rho_{2}
        & = & {\rm Tr}_{A}\, |2\rangle\langle 2| .
\end{eqnarray*}
We define (as in \cite{jozsa}) the fidelity $F(\rho_{1},\rho_{2})$ by
\begin{equation}
        F(\rho_{1},\rho_{2}) = \max |\langle 1|2\rangle|^{2}\;,
\end{equation}
where the maximum is taken over all purifications $|1\rangle$ of
$\rho_{1}$ and $|2\rangle$ of $\rho_{2}$.  Thus, the fidelity is the
largest squared inner product between purifications of two density
operators.  This definition provides a generalization to mixed states
of the natural squared inner product measure of fidelity for pure
states.

Basic properties of this notion of fidelity are described in detail in
\cite{jozsa} and we note the following.
\begin{itemize}
\item $ 0\leq F(\rho_{1},\rho_{2})\leq 1$ and $F(\rho_{1},\rho_{2})=1$
if and only if $\rho_1 = \rho_2 $.
\item $F(\rho_{1},\rho_{2}) = F(\rho_{2},\rho_{1})$.
\item If one of the states $\rho_{1}$ is a projection $\pi_{1}$, i.e.,
a {\em pure\/} state, then we have the more direct expression
\begin{displaymath}
        F(\pi_{1},\rho_{2}) = {\rm Tr}\, \pi_{1} \rho_{2} \;.
\end{displaymath}
(A general expression for arbitrary mixed states is given in
\cite{jozsa} but this is not required in the present work.)
\item In defining the fidelity for mixed states, 
it is sufficient to fix
any one of the purifications $|1\rangle$ of $\rho_{1}$ and take the
maximum of $|\langle 1|2\rangle|^{2}$ over arbitrary purifications
$|2\rangle$ of $\rho_{2}$.
\end{itemize}
We can extend the definition of fidelity from normalized states to
subnormal\-ized states (in which ${\rm Tr}\, \rho_{1} < 1$) in an
obvious way, by requiring that the purifications have the same
normalization: $\langle 1|1\rangle = {\rm Tr}\,\rho_{1}$.

We now establish a useful inequality for fidelity.  Let $\rho_{1}$,
$\rho_{2}$, and $\rho_{3}$ be states, and let $F_{12} =
F(\rho_{1},\rho_{2})$, etc.  We will require that ${\rm Tr}\,\rho_{3}
= 1$, but $\rho_{1}$ and $\rho_{2}$ may be subnormalized.  Then
\begin{equation}
F_{13} \leq F_{23} + 2\!\left(1-\sqrt{F_{12}}\,\right) + 2 \sqrt{ 2 }
               \sqrt{ F_{23}\left(1 - \sqrt{F_{12}}\,\right) }\;.
               \label{fidineq}
\end{equation}
This implies that if $F_{12}$ is close to unity and $F_{23}$ is close
to zero, then $F_{13}$ must also close to zero.

        The proof is not difficult.  We construct purifications for
our states with these properties:
\begin{itemize}
        \item All inner products ($\langle 1|2\rangle$, etc.) are real
                and nonnegative, \item $F_{12} = \langle
                1|2\rangle^{2}$, \item $F_{13} = \langle
                1|3\rangle^{2}$.
\end{itemize}
This can be done by the following procedure.  We fix $|1\rangle$ and
choose $|2\rangle$ and $|3\rangle$ so that $F_{12} = |\langle
1|2\rangle|^{2}$ and $F_{13} = | \langle 1|3\rangle |^{2}$.  Next we
adjust the phases of $|1\rangle$, $|2\rangle$, and $|3\rangle$ to
satisfy the first condition.  Clearly, $F_{23} \geq
\langle 2|3\rangle^{2}$.

Let $|x\rangle = |2\rangle - |1\rangle$.  Then
\begin{eqnarray*}
\langle x|x\rangle
&=& \langle 1|1\rangle + 
\langle 2|2\rangle- 2 \langle 1|2\rangle \\ &
                \leq & 2\! \left( 1 - \sqrt{F_{12}}\,\right) \;,
\end{eqnarray*}
because $\rho_{1}$ and $\rho_{2}$ may be subnormalized.
Furthermore,
\begin{eqnarray*}
        \sqrt{F_{13}} & = & \langle 1|3\rangle \\ & = & \langle
        2|3\rangle - \langle x|3\rangle \\ & \leq & \sqrt{F_{23}}+|
        \langle x|3\rangle | \\ & \leq & \sqrt{F_{23}}+\sqrt{\langle
        x|x\rangle} \\ & \leq & \sqrt{F_{23}}+\sqrt{2\!\left( 1 -
        \sqrt{F_{12}}\,\right) }\;.
\end{eqnarray*}
Thus,
\begin{displaymath}
        F_{13} \leq F_{23} + 2\!\left(1-\sqrt{F_{12}}\,\right) + 2
             \sqrt{ 2 } \sqrt{F_{23}\left(1 - \sqrt{F_{12}}
\,\right) }
\end{displaymath}
as we wished to prove.

We note in passing that, if we relax the condition that ${\rm Tr}\,
\rho_{3} = 1$, we arrive at the more general inequality for
subnormalized states:
\begin{eqnarray*}
F_{13} &\le& F_{23}\,+\,2\,{\rm
Tr}\,\rho_{3}\left(1-\sqrt{F_{12}}\,\right)\\ &&\phantom{F_{23}\,} 
+ 2
\sqrt{ 2\,{\rm Tr}\, \rho_{3}} \, \sqrt{F_{23}
\left(1 - \sqrt{F_{12}}\,\right) }\;.
\end{eqnarray*}

\section{Channel size and fidelity}

The ``size'' of the channel system $C$ is specified by the dimension
$d$ of the Hilbert space describing $C$.  If $C$ is composed of $M$
qubits, then $d = 2^{M}$.  This means that in the process
\begin{displaymath}
        \pi_{i}  \longrightarrow
        W_{i}  \longrightarrow  w_{i} \;.
\end{displaymath}
the channel states $W_{i}$ are operators on a $d$-dimensional Hilbert
space.  For convenience, we will imagine that the $W_{i}$ actually act
on a $d$-dimensional subspace of the $n$-dimensional Hilbert space
describing the system $Q$.  (We could always modify our decoding
procedure so that the channel states were first unitarily moved into
the output system $Q$ and then subjected to a more general decoding
process.  The $W_{i}$ states would then be the unitary images of the
channel states in $Q$'s Hilbert space.)

We are now ready to state our result.  Imagine that an ensemble of
pure states of $Q$ (in which the state $\pi_{i}$ appears with
probability $p_{i}$) is described by a density operator $\rho =
\sum_{i} p_{i} \pi_{i}$.  Let $\lambda_{i}$ be the eigenvalues of
$\rho$, listed in descending order (so that $\lambda_{1} \geq \ldots
\geq \lambda_{n}$), and let $|\lambda_{i}\rangle$ be associated
eigenvectors.
\begin{quote}
        {\bf Fidelity lemma:} Suppose the dimension of the Hilbert
        space for the channel is $d$, and write \begin{displaymath}
        \sum_{i=1}^{d} \lambda_{i} = \eta \;.  \end{displaymath} Then,
        for any encoding and decoding procedures, $\overline{F} < 6
        \eta$.
\end{quote}

To prove this lemma, we first note that
\begin{displaymath}
        \eta = \sum_{i=1}^{d} \lambda_{i} \geq d \lambda_{d+1}
\end{displaymath}
so that $\lambda_{d+1} \leq \eta / d$.  Now we construct a projection
operator
\begin{displaymath}
        \Lambda = \sum_{i = d+1}^{n}
|\lambda_{i}\rangle\langle\lambda_{i}|\;,
\end{displaymath}
which is the projection onto the subspace spanned by the eigenvectors
corresponding to the $n-d$
smallest eigenvalues of $\rho$.  We use $\Lambda$ to project the input
states $\pi_{i}$ into (subnormalized) states $\tilde{\pi}_{i}$:
\begin{eqnarray*}
        \tilde{\pi}_{i} & = & \Lambda \pi_{i} \Lambda
        \phantom{\sum_i}\\ \tilde{\rho} & = & \sum_{i} p_{i}
        \tilde{\pi}_{i} = \Lambda \rho \Lambda \;.
\end{eqnarray*}
The largest eigenvalue of $\tilde{\rho}$ is just $\lambda_{d+1}$.

Our plan is as follows.  (For heuristic purposes and later
application, we have in mind a situation with $\eta$ small.)  First,
we will show that the original input states $\pi_{i}$ are, on average,
close to the projected states $\tilde{\pi}_{i}$.  Then we will show
that the average of $F(\tilde{\pi}_{i},w_{i})$ is small for all
possible coding/decoding schemes.  Using the fidelity inequality in
equation~\ref{fidineq} above, we will conclude that the average of
$F(\pi_{i},w_{i})$ must therefore be small.  The qualitative phrases
``close to'' and ``small'' will be quantified by the value of $\eta$.

Anticipating somewhat, we first find a lower bound for the average of
the {\em square root\/} of $F(\pi_{i},\tilde{\pi}_{i})$.  Recall that
$\pi_{i} = |a_{i}\rangle\langle a_{i}|$.
\begin{eqnarray}
\sum_{i} p_{i} \sqrt{F(\pi_{i},\tilde{\pi}_{i})}
        & = & \sum_{i} p_{i} \sqrt{ {\rm
                Tr}\,\pi_{i}\Lambda\pi_{i}\Lambda } \nonumber \\ & = &
                \sum_{i} p_{i} \sqrt{ \langle a_{i}| \Lambda
                |a_{i}\rangle\langle a_{i}| \Lambda |a_{i}\rangle}
                \nonumber \\ & = & \sum_{i} p_{i} \langle a_{i}|
                \Lambda |a_{i}\rangle\nonumber\\ & = & {\rm Tr}\, \rho
                \Lambda \nonumber \\ & = & 1 - \eta \;.  \label{yineq}
\end{eqnarray}

We wish the decoding procedure to be as general as possible.
Therefore we only require that the procedure be specifiable
independently of the state $W_{i}$ to which it is applied, and that it
is an allowable quantum dynamical evolution.  The most general
dynamical evolution possible in quantum mechanics is a completely
positive map on the space of density operators
\cite{kraus}.  Such a map can always be modeled by a unitary interaction
between the system $Q$ and an ancilla system $A$ (initially in some
standard pure state $|\phi_{0}\rangle$), after which $A$ is discarded.
We can therefore write
\begin{equation}
w_{i}={\rm Tr}_{A}\,U(
W_{i}\otimes|\phi_{0}\rangle\langle\phi_{0}|)U^{\dagger}
\end{equation}
for some unspecified unitary $U$.

We can use this general form to find an upper bound for the average of
$F(\tilde{\pi}_{i},w_{i})$.
Note that, although $\tilde{\pi}_{i}$ is subnormalized, it is still an
operator of rank 1, and thus we can write the fidelity as
${\rm Tr}\, \tilde{\pi}_{i} w_{i}$.
Let $\Gamma_{d}$ be the projection onto the $d$-dimensional subspace
occupied by the channel states $W_{i}$.  Then, writing the trace over
the $Q$ Hilbert space as ${\rm Tr}_{Q}$, etc.,
\begin{eqnarray*}
F(\tilde{\pi}_{i},w_{i})
&=&
\sum_{i} p_{i}\, {\rm Tr}_{Q}\, \tilde{\pi}_{i} 
\Bigl(  {\rm Tr}_{A} U (
W_{i} \otimes |\phi_{0}\rangle\langle\phi_{0}| ) 
U^{\dagger} \Bigr) \\
&=&
\sum_{i} p_{i}\, {\rm Tr}_{QA}\, ( \tilde{\pi}_{i} \otimes 1_{A} ) U
(W_{i} \otimes |\phi_{0}\rangle\langle\phi_{0}| ) U^{\dagger} \\
&\leq&
\sum_{i} p_{i}\, {\rm Tr}_{QA}\, ( \tilde{\pi}_{i} \otimes 1_{A} ) U 
(\Gamma_{d} \otimes |\phi_{0}\rangle\langle\phi_{0}| ) U^{\dagger} \\
&=&
{\rm Tr}_{QA}\, ( \tilde{\rho} \otimes 1_{A} ) U (\Gamma_{d} \otimes 
|\phi_{0}\rangle\langle\phi_{0}| ) U^{\dagger}\;.
\end{eqnarray*}
Now, every eigenvalue of $\tilde{\rho} \otimes 1_{A}$ is an eigenvalue
of $\tilde{\rho}$.  Furthermore, the operator $U (\Gamma_{d} \otimes
|\phi_{0}\rangle\langle\phi_{0}| ) U^{\dagger}$ is a projection onto a
$d$-dimensional subspace.  The trace will therefore be less than or
equal to the sum of the $d$ largest eigenvalues of $\tilde{\rho}
\otimes 1_{A}$, which in turn can be no larger than $d \,
\lambda_{d+1}$:
\begin{eqnarray}
\sum_{i} p_{i}\, {\rm Tr}_{Q}\, \tilde{\pi}_{i} w_{i}
         & \le & d \, \lambda_{d+1} \nonumber \\ & \le & d \left (
         \frac{\eta}{d} \right) = \eta \;. \label{xineq}
\end{eqnarray}

We now find an upper bound for $\overline{F}$ by applying the 
fidelity
inequality in equation~\ref{fidineq} to each term in the average:
\begin{eqnarray*}
F(\pi_{i},w_{i})\,
&\leq&
\,\underbrace{F(\tilde{\pi}_{i},w_{i})}_{X_{i}}
\,+\,\underbrace{2\!\left(1-\sqrt{F(\pi_{i},
\tilde{\pi}_{i})}\,\right)}_{Y_{i}}
\\
&&\phantom{F(\tilde{\pi}}+\,
\underbrace{  2 \sqrt{ 2 F(\tilde{\pi_{i}},w_{i})\!
\left(1 - \sqrt{F(\pi_i,\tilde{\pi}_i ) }\,\right) } }_{Z_i}\;.
\end{eqnarray*}
We will bound the averages $\overline{X}$, $\overline{Y}$, and
$\overline{Z}$ separately.

We have already bounded $\overline{X}$ in equation~\ref{xineq}.
\begin{displaymath}
        \overline{X} = \sum_{i} p_{i} X_{i} =
        \sum_{i} p_{i}\, {\rm Tr}_{Q} \tilde{\pi}_{i} w_{i} \le 
\eta\;.
\end{displaymath}
Similarly, the bound for $\overline{Y}$ follows from equation~
\ref{yineq}.
\begin{eqnarray*}
        \overline{Y} & = & \sum_{i} p_{i} Y_{i} \\
                & = & 2\! \left ( 1 -
                        \sum_{i} p_{i} \sqrt{F(\pi_{i},
\tilde{\pi}_{i})}
                        \right ) \\
                & = & 2 \eta \;.
\end{eqnarray*}
To find an upper bound for $\overline{Z}$, we use these two results
together with the Schwarz inequality:
\begin{eqnarray*}
        \overline{Z} & = & \sum_{i} p_{i} Z_{i} \\
                & = & 2 \sum_{i} p_{i} \sqrt{X_{i} Y_{i}} \\
                & \leq & 2 \sqrt{\sum_{i} p_{i} X_{i}}
                           \sqrt{\sum_{j} p_{j} Y_{j}} \\
                & \le & 2 \sqrt{2} \, \eta \;.
\end{eqnarray*}
Therefore,
\begin{eqnarray*}
\overline{F} & = & \overline{X} + \overline{Y} + \overline{Z} \\
        & \le & \eta + 2 \eta + 2 \sqrt{2} \, \eta \, < \, 6 \eta \;,
\end{eqnarray*}
which is what we wished to establish.

We point out once again that no assumption has been made about the
encoding procedure $\pi_{i} \rightarrow W_{i}$.  This may be
completely arbitrary.  We do not require that it be accomplished by a
process that is ``blind'' to the input state $\pi_{i}$, that is, by a
completely positive map.  This means that we are allowing Alice to be
completely cognizant of the identity of the input she is representing
in the channel, even though it may be one of a nonorthogonal (and
hence imperfectly distinguishable) set.

We note finally that the bound $\overline{F} < 6 \eta$ is quite likely
to be loose.  For example, in \cite{schu} and \cite{js94}, where the
decoding scheme was assumed to be unitary, a bound of $\overline{F}
\le \eta$ was derived.  This bound for unitary decoding is achieved by
a very natural coding/decoding scheme---$W_i$ is the renormalized
projection of $\pi_i$ into the subspace corresponding to $\rho$'s
largest $d$ eigenvalues and the unitary decoding is just the identity.
Denoting the projector onto this subspace by $\Gamma_d$, the fidelity
may be written (taking the sum to exclude $i$ such that $\pi_i$ are
orthogonal to $\Gamma_d$, which make zero contribution to average
fidelity however they are encoded):
\begin{eqnarray*}
     \overline{F} & = & \sum_i p_i {\rm
             Tr}\!\left(\pi_i\frac{\Gamma_d\pi_i\Gamma_d} {{\rm
             Tr}\,\pi_i\Gamma_d}\right) \\ & = & \sum_i p_i
             \frac{\langle a_i|\Gamma_d| a_i \rangle \langle
             a_i|\Gamma_d| a_i \rangle} {\langle a_i|\Gamma_d| a_i
             \rangle} \\ & = & \sum_i p_i \langle a_i|\Gamma_d| a_i
             \rangle \\ & = & {\rm Tr}\,\rho\Gamma_d \\ & = & \eta\;.
\end{eqnarray*}
Nevertheless the bound of $6\eta$ suffices for proving the converse of
the quantum noiseless coding theorem.

\section{Quantum coding}

Suppose the input state $\pi_{i}$ of $Q$ occurs with probability
$p_{i}$, so that the ensemble of inputs is described by $\rho =
\sum_{i} p_{i} \pi_{i}$, as above.  Further suppose that a long
sequence of $N$ such inputs, generated independently, is available.
The ensemble of $N$-sequences of input states is then described by
\begin{displaymath}
        \rho^{N} = \overbrace{\rho \otimes \cdots \otimes \rho}^{N} \;.
\end{displaymath}
For sufficiently large $N$, the structure of $\rho^{N}$ is
characterized by a {\em typical subspace\/} ${\cal T}_{N}$
\cite{schu,js94}.

The typical subspace may be described as follows.  Fix $\epsilon,
\delta > 0$.  Then for sufficiently large $N$, there exists a subspace
${\cal T}_{N}$ spanned by eigenstates of $\rho^{N}$ such that
\begin{itemize}
        \item If $\Pi$ is the projection onto ${\cal T}_{N}$, then
              \begin{displaymath} {\rm Tr}\,\Pi \rho^{N} \Pi > 1 -
              \epsilon\;.  \end{displaymath} \item If
              $|\lambda\rangle$ is an eigenstate of $\rho^{N}$ with
              eigenvalue $\lambda$, and $|\lambda\rangle \in {\cal
              T}_{N}$, then \begin{displaymath} 2^{-N(S(\rho) +
              \delta)} < \lambda < 2^{-N(S(\rho) - \delta)}\;.
              \end{displaymath}
\end{itemize}

Now suppose that a sequence of $N$ inputs is encoded somehow into a
set of qubits, so that $S(\rho) - 2\delta$ qubits are used per input.
The Hilbert space describing the channel of $N(S(\rho) - 2\delta)$
qubits will have dimension $d = 2^{N(S(\rho) - 2\delta)}$.  The
channel states are used in some decoding procedure to produce an
output state of $N$ copies of $Q$.

According to our fidelity lemma, we can bound the fidelity of this
process by calculating the sum of the largest $d$ eigenvalues of
$\rho^{N}$.  We will denote this by $\Sigma_{d}$.  This sum must
certainly be smaller than the sum of all of the eigenvalues outside
the typical subspace ${\cal T}_{N}$ plus $d$ times the largest
eigenvalue inside ${\cal T}_{N}$.  That is,
\begin{eqnarray*}
\Sigma_{d}  & < &       \epsilon  +  d \, 2^{-N(S(\rho) - \delta)} \\
            & = &       \epsilon  +  2^{N(S(\rho) - 2\delta)}
                                        2^{-N(S(\rho) - \delta)} \\
            & = &       \epsilon + 2^{-N \delta} \;.
\end{eqnarray*}
For sufficiently large $N$, $\Sigma_{d} < 2 \epsilon$.  Thus, by our
fidelity lemma, $\overline{F} < 12 \epsilon$.  Letting $\delta =
\delta'/2$ and $\epsilon = \epsilon'/12$, we find that if $S(\rho) -
\delta'$ qubits are available per input, then for sufficiently large
$N$ the average fidelity $\overline{F} < \epsilon'$.  This establishes
the converse to the quantum noiseless coding theorem for the most
general sort of coding and decoding schemes.

\section{Appendix}

We demonstrate here by explicit example that decoding schemes more
general than the set of unitary ones can be of some benefit in
situations of nonoptimal coding.

Consider three signal states $|a_0\rangle$, $|a_1\rangle$, and
$|a_2\rangle$ which are all real positive linear combinations of three
fixed orthonormal vectors, so that we may picture them as vectors in
the positive octant of $\mbox{I\kern-.2emR}^3$.  The states form three
edges of a regular tetrahedron with the origin as their common vertex,
and thus are all $60^\circ$ apart.  The states $|a_0\rangle$ and
$|a_1\rangle$, in particular, are assumed to be in the positive
quadrant of the $x$-$y$ plane, each vector having an angle of
$15^\circ$ between itself and the nearest axis.  The prior
probabilities for the signal states are $.49$, $.49$, and $.02$,
respectively.  The encoding scheme associates the orthogonal
projectors $W_0$ and $W_1$ onto the $x$ and $y$ axes, respectively,
with the states $|a_0\rangle$ and $|a_1\rangle$.  It associates the
density matrix $$ W_2= \frac{1}{2}|a_0\rangle\langle a_0| +
\frac{1}{2}|a_1\rangle\langle a_1|\;, $$ corresponding to an equal
mixture of $|a_0\rangle$ and $|a_1\rangle$, with the state
$|a_2\rangle$.  Note that the set of encoded states has a
two-dimensional support, i.e., a support smaller than that containing
the signal states.

Because the signal state $|a_2\rangle$ has such a small prior
probability, the symmetry of this encoding should make it clear that
the best unitary decoding scheme will be only slightly different from
{\em not\/} decoding at all.  (Actually, detailed calculation
demonstrates that the optimal unitary decoding is to rotate the
encoded states by $0.791^\circ$ toward $|a_2\rangle$, but this only
changes the average fidelity in the fourth significant figure.)
Making this approximation, the average fidelity for this decoding
scheme is $$
\overline{F}\,=\,2\times .49\times\cos^2 15^\circ\;+\;
.02\times\cos^2 60^\circ\,=\,.919\;.
$$

However there exists a simple nonunitary decoding scheme that achieves
a better fidelity than this.  Since some of the signals are encoded in
orthogonal alternatives, it is plausible that a decoding device can
use a measurement to gather information about the signal and use that
information to produce decoded states that are closer, on average, to
the originals.  In particular, the decoding device can do the
following.  It first measures the observable corresponding to the
$x$-$y$ axis.  If the outcome is $x$, it outputs the state
$w_0=\pi_0$; if the outcome is $y$, it outputs the state $w_1=\pi_1$.
Thus in the cases that $Q$ was actually prepared in $|a_0\rangle$ or
$|a_1\rangle$, the transmissions will have perfect fidelity.  In the
case that $|a_2\rangle$ was the actual signal state, the fidelity of
the transmission will still be $\cos^2 60^\circ=.25$.  Therefore the
average fidelity for this nonunitary decoding scheme is
$\overline{F}=.985$, and this certainly beats the unitary scheme.

This simple example demonstrates that in some cases
involving {\em particular nonoptimal\/}
encoding schemes, it is possible for nonunitary
decoding to increase the fidelity of a quantum channel.
Nevertheless the converse of the quantum noiseless theorem implies that
nonunitary decodings provide no asymptotic advantage
over unitary decoding schemes
in the problem of minimizing of channel resources over all possible
coding/decoding schemes.

\section*{Acknowledgments}

We thank Carlton Caves, Michael Nielsen, and Michael Westmoreland
for many helpful discussions during the course of this work.
BS thanks the Theoretical Astrophysics group (T-6) at Los Alamos
National Laboratory for hospitality and support during 1995--96.
This work was supported in part by the Office of Naval Research
(Grant No. N00014-93-1-0116).



\begin{references}

\bibitem[\dagger]{FuchsAddress}
Permanent Address: D\'{e}partement IRO, Universit\'{e} de
Montr\'{e}al, C.~P.  6128, Succursale centre-ville, Montr\'eal,
Qu\'ebec, Canada H3C 3J7.

\bibitem[*]{JozsaAddress}
Permanent Address: School of Mathematics and Statistics, University of
Plymouth, Drake Circus, Plymouth, Devon PL4 8AA, England.

\bibitem[\ddagger]{SchumacherAddress}
Permanent Address: Department of Physics, Kenyon College, Gambier,
Ohio 43022.

\bibitem{chb}  C. H. Bennett, {\em Physics Today\/} 
{\bf 48}, 24--30 (1995).

\bibitem{schu}  B. Schumacher, {\em Phys.\ Rev.\ A\/} 
{\bf 51}, 2738--2747 
        (1995).

\bibitem{js94}  R. Jozsa and B. Schumacher, {\em J. Mod.\ Optics\/} 
{\bf 41},
        2343--2349 (1994).

\bibitem{Bennett84}  C. H. Bennett and G. Brassard, in 
{\em Proceedings of
        IEEE International Conference on 
Computers, Systems and Signal
        Processing, Bangalore, India\/} (IEEE, New York, 1984)
        175--179.

\bibitem{Bennett92} C. H. Bennett, {\em Phys. Rev. Lett.} 
{\bf 68}, 3121--3124
        (1992).

\bibitem{DiVincenzo95} D. P. DiVincenzo, {\em Science\/} 
{\bf 279}, 255--261
        (1995).

\bibitem{Wootters82} W.~K. Wootters and W.~H. Zurek,  {\em Nature\/} 
{\bf 299},
        802--803, (1982).

\bibitem{asher} A. Peres, {\em Phys.\ Lett.\ A\/}, 
{\bf 128}, 19 (1988).

\bibitem{GP} C. A. Fuchs and A. Peres, ``Quantum State Disturbance 
        vs.\ Information Gain: Uncertainty Relations for Quantum
        Information,'' to appear in {\em Phys.\ Rev.\ A\/} (1996).

\bibitem{jozsa}  R. Jozsa, {\em J. Mod.\ Optics\/} 
{\bf 41}, 2315--2323 (1994).

\bibitem{kraus}  K. Hellwig and K. Kraus, {\em Comm.\ Math.\ Phys.}
        {\bf 16}, 142--147 (1970); K.~Kraus, 
{\em States, Effects, and
        Operations: Fundamental Notions of Quantum Theory\/}
        (Springer, Berlin, 1983).

\end{references}
\end{document}